\documentclass[aps,prb,twocolumn,superscriptaddress,eqsecnum]{revtex4}
\usepackage{amsmath}
\usepackage{color}
\usepackage{graphicx}
\usepackage{subfigure}
\usepackage{hyperref}
\usepackage{bm}
\usepackage{amsfonts}

\def\be{\begin{equation}}       \def\ee{\end{equation}}
\def\bea{\begin{eqnarray}}      \def\eea{\end{eqnarray}}
\def\ba{\begin{array} }
\def\ea{\end{array} }
\def\bnum{\begin{enumerate} }
\def\enum{\end{enumerate}}

\def\=>{\Rightarrow}
\def\>{\rightarrow}

\def\d0{\Delta_{0}}

\def\C60{A$_x$C$_{60}$}

\begin{document}
\title{\bf 
Fragile superconductivity in the presence of weakly disordered charge density waves
}
\author{Yue Yu and S. A. Kivelson}
\affiliation{Department of Physics, Stanford University, Stanford, CA 94305, USA}

\date{\today }

\begin{abstract}
When superconducting (SC) and charge-density wave (CDW) orders compete, novel low temperature behaviors can result.  From an analysis of the Landau-Ginzburg-Wilson theory of competing orders, we demonstrate  the generic occurrence of a ``fragile'' SC phase at low temperatures and high fields  in the presence of weak disorder.  Here, the SC order is largely concentrated in the vicinity of dilute dislocations in the CDW order, leading to transition temperatures and critical currents that are parametrically smaller than those characterizing the zero field SC phase. This may provide the outline of an explanation of the recently discovered ``resilient'' superconducting phase at high fields in underdoped YBa$_2$Cu$_3$O$_{6+\delta}$.
\end{abstract}

\maketitle

\section{Introduction}
There are a variety of microscopic circumstances in which correlated electronic materials exhibit comparably strong, and macroscopically competing tendencies toward superconducting (SC) and incommensurate charge-density wave (CDW) orders.\cite{CDW}  A particularly interesting aspect of the interplay between the two orders is the role of topological defects -- vortices in the SC, which can be induced by the application of an external magnetic field, $H$, and dislocations in the CDW, which are produced by quenched disorder.  In cases in which a subdominant order is nearly degenerate with the dominant order, the subdominant order can be stabilized in the neighbor of a topological defect where the dominant order vanishes. The possibility of one or another form of density wave order  in a vortex core ``halo'' has been the subject of considerable theoretical and experimental interest, especially in the context of the underdoped cuprate high temperature superconductors.  Here we explore some qualitative new physics that arises in regimes where CDW order is dominant, and in which SC arises in the neighborhood of CDW dislocations.

\subsection{Topological defects and  halos}
\label{halo}

In systems with complex phase diagrams, there is often a range of temperatures, $T$, in which -- in the absence of competition -- two distinct symmetries would be spontaneously broken, but where the competition between the two orders is strong enough that only the dominant order parameter develops a non-zero expectation value;  the subdominant order is quenched by the competition.  This state of affairs pertains when the repulsion between the two orders, which corresponds to the term $\gamma$ in the Landau-Ginzburg free energy in Eq. \ref{H}, exceeds a critical strength, $\gamma_{c1}$.  (See Figs. \ref{1a} and \ref{1b}.)  However, under some circumstances (that we will derive explicitly), the subdominant order emerges in a ``halo'' of finite extent surrounding any topological defect in the primary order.
Two features of the system are required for this to occur:  
\begin{itemize}
\item Firstly, the magnitude of the primary order must vanish -- or at least must be substantially suppressed - in the core of the defect.  This is generic in the vortex core of an order parameter with XY symmetry, or a domain wall of an order parameter with Ising symmetry.  It would not be expected, for instance, in a skyrmion of an order parameter with Heisenberg symmetry.   

\item Secondly, the elastic ``stiffness'' of the secondary order parameter, $\kappa$, must be sufficiently small, so that the gain in condensation energy from having the secondary order expressed in the defect core exceeds the elastic cost of having a spatially varying order parameter. (Note, the critical value of $\kappa$ depends, among other things, on how close is the balance is between the primary and secondary ordering tendencies.)
\end{itemize}

Under these circumstances, there exists a second critical value, $\gamma_{c2}(\kappa) > \gamma_{c1}$ such that for $\gamma_{c2} > \gamma > \gamma_{c1}$, there occurs a finite halo region about any topological defect of the primary order parameter in which the secondary order parameter has a non-vanishing amplitude.  There are many examples of this basic physics that have been discussed.  Superconducting cosmic strings\cite{cosmicstrings1,cosmicstrings2} are an example, in which the superconducting order is the secondary order that appears associated with vortices in a dominant cosmic condensate.  Subdominant CDW, spin-density wave (SDW), or nematic orders  arising in halos about magnetic field induced vortices in a dominantly superconducting order have been theoretically discussed\cite{Zhang1089,PhysRevLett.79.2871,PhysRevB.63.224517,PhysRevLett.87.067202,PhysRevB.66.094501,PhysRevB.66.144516,kivelson-2002a,PhysRevB.91.104512,PhysRevB.97.174510,PhysRevB.97.174511}, and  experimentally observed\cite{lake-2001,lake-2005,hoffman-2002a,edkins-2018}, in various cuprate superconductors.  Some evidence has been presented that in certain Fe-pnictides, superconducting order can exist in a narrow range of $T$ above the bulk $T_c$ in a region about a structural twin boundary - i.e. Ising domain walls of a nematic order parameter.\cite{PhysRevB.81.184513}

In the present paper, we will focus on the case of competition between CDW and SC order.  Where the SC order is dominant, the topological defects in question are the familiar vortices already mentioned, and the halo is then a region with local CDW order.  Here, at low $T$ in a strongly type II superconductor, the density of vortices is controlled by an applied magnetic field, $H$, and a low density of vortices can be introduced by the application of a small field with $H_{c1} < H \ll H_{c2}$.  Conversely, where the CDW order is dominant, the topological defects are dislocations.  Because there is an emergent XY symmetry associated with an incommensurate CDW, at a formal level these defects are precisely dual to the superconducting vortices, with the role of CDW and SC orders interchanged.  Now, the halo is a region of local SC order in the vicinity of a dislocation core.  No precise dual relation exists in the presence of a magnetic field, which for all practical purposes couples mimimally only to the SC order.  However,  quenched disorder  plays a somewhat analogous role;  in a quasi-2D system, weak disorder (greater than a parametrically small value that vanishes as interplane couplings tend to 0) produces dilute, randomly pinned dislocations in the CDW order \cite{Tissier-2006}.  (See also, \onlinecite{Imry-Ma-1975,Aizenman-1989,Giamarchi-1994,Gingras-1996}.)

\subsection{Halos and Long-Range-Order}
\label{fragileLRO}

Until now, we have focussed on local (or mean-field) considerations.  An isolated vortex or dislocation is a 1D object (or a 0D object in a 2D system), so an isolated vortex halo cannot give rise to a broken symmetry. \cite{kivelson-2002a}.  Unless inter-halo interactions are taken into account, thermal fluctuations destroy any long-range-order (LRO) that one would have associated with the order parameter halo.  Thus, LRO (if it occurs) is triggered by the coupling between  halos.  As a result, for  a system of dilute halos, there are typically two parametrically distinct characteristic temperatures:  a relatively high crossover temperature $T_{hal}$ below which the halos are locally well formed and a lower critical temperature, $T_{c}$, at which LRO of the subdominant order parameter onsets.  Given that the effective exchange couplings between well separated halos have random phase, the nature of the LRO that results is extremely complicated, and not well understood theoretically -- the ordering problem is some form of ``XY-gauge glass problem.''\cite{Katzgraber-2002}  In the LGW we have considered, these couplings fall exponentially with separation between halos.  However, as $T\to 0$, while the basic structure of the state remains unchanged, the details of these effective couplings are expected to be increasingly altered by the existence of gapless electronic quasiprticles which typically mediate interactions that fall with an inverse power of separation.

For the problem at hand, we will treat these issues by explicit solution of a simple effective field theory.  However, to develop a general intuition,  we can estimate $T_{c}$ as follows:  1) Compute the order parameter susceptibility of an isolated  halo, $\chi(T)$.  For instance, the susceptibility associated with a CDW halo living along a vortex line is that of an appropriate classical 1D XY model, $\chi(T) \sim 1/\sqrt{T^\star T}$ (where $T^\star$ is the mean-field $T_c$).  2)  Compute the effective coupling between neighboring halos, $J(R)$, which naturally depends strongly on the distance, $R$, between halos.  In the simple Landau-Ginzburg effective field theories we will analyze, $J(R) \sim J_0 \exp[ -R/\ell]$, where $\ell$ is an appropriate correlation length;  in more microscopically realistic treatments of  metallic systems at low $T$, this dependence can be much more complicated.\cite{FiigelmanLarkinSkvortsov,OretoKivSp}   $T_{c}$ can be estimated from the implicit solution of the equation $\chi(T_{c}) J(R) \sim 1$. 
\\

\subsection{Plan of the paper}
\label{plan}

Before launching into specific calculations, in Sec. \ref{Qualitative} we present some representative phase diagrams that can emerge as a consequence of the competition between CDW and SC order.  Next, in
 Sec. \ref{Model}, we define a model classical effective field theory with fields corresponding to unidirectional CDW and SC order.  In Sec. \ref{meanfield} we treat this model in the context of  Landau-Ginzburg theory, meaning that we treat the model as an effective free energy (with coefficients that are assumed to depend smoothly on  $T$) and we solve for the field configurations that minimize it.  
 
In order to explore the effects of order parameter fluctuations more seriously, in Secs. \ref{fluct} and \ref{mc} we treat the same model as an effective Hamiltonian (i.e. in the context of Landau-Ginzburg-Wilson theory), with fixed (temperature independent) parameters.  Specifically, in Sec. \ref{fluct} we treat the fluctuations of the fields approximately using Feynman's variational approach\cite{feynman} (which is exact in a suitable large $N$ limit of the same model\cite{sachdev}) and in Sec. \ref{mc} we treat the fluctuations exactly (numerically) using classical Monte-Carlo methods.  

 Finally, in Sec.~\ref{htc} we generalize the discussion slightly to the case of a tetragonal system (where the CDW order can have its ordering vector in one of two symmetry related directions) in order to obtain a phase diagram that is suggestive of observed behaviors of the cuprate high temperature superconductors.  We then comment on insights one can obtain concerning existing observations in the cuprates that can be qualitatively well understood in terms of the present considerations. In particular, in the interest of brevity, we focus on YBCO with doped hole density roughly in the range 0.1 to 0.14, which we identify as a regime in which the principle features of the phase diagram  are  a consequence of the fierce competition between SC and CDW orders.
While in this paper, we have considered only problems in which there is a close competition between CDW and SC order, very similar considerations apply to other cases in which multiple orders are intertwined.\cite{varma-1997,kivelson-1998,CASTELLANI19981694,chakravarty-2001c,RevModPhys.75.1201,RevModPhys.75.913,intertwined,DavisLee,PhysRevB.78.174529,daSilvaNeto393,PhysRevB.90.035149,PhysRevB.89.075129,PhysRevB.88.020506,PhysRevB.82.075128}

\section{Qualitative phase diagrams}
\label{Qualitative}

\begin{figure*}[t]
\subfigure[]
{\includegraphics[scale=0.3]{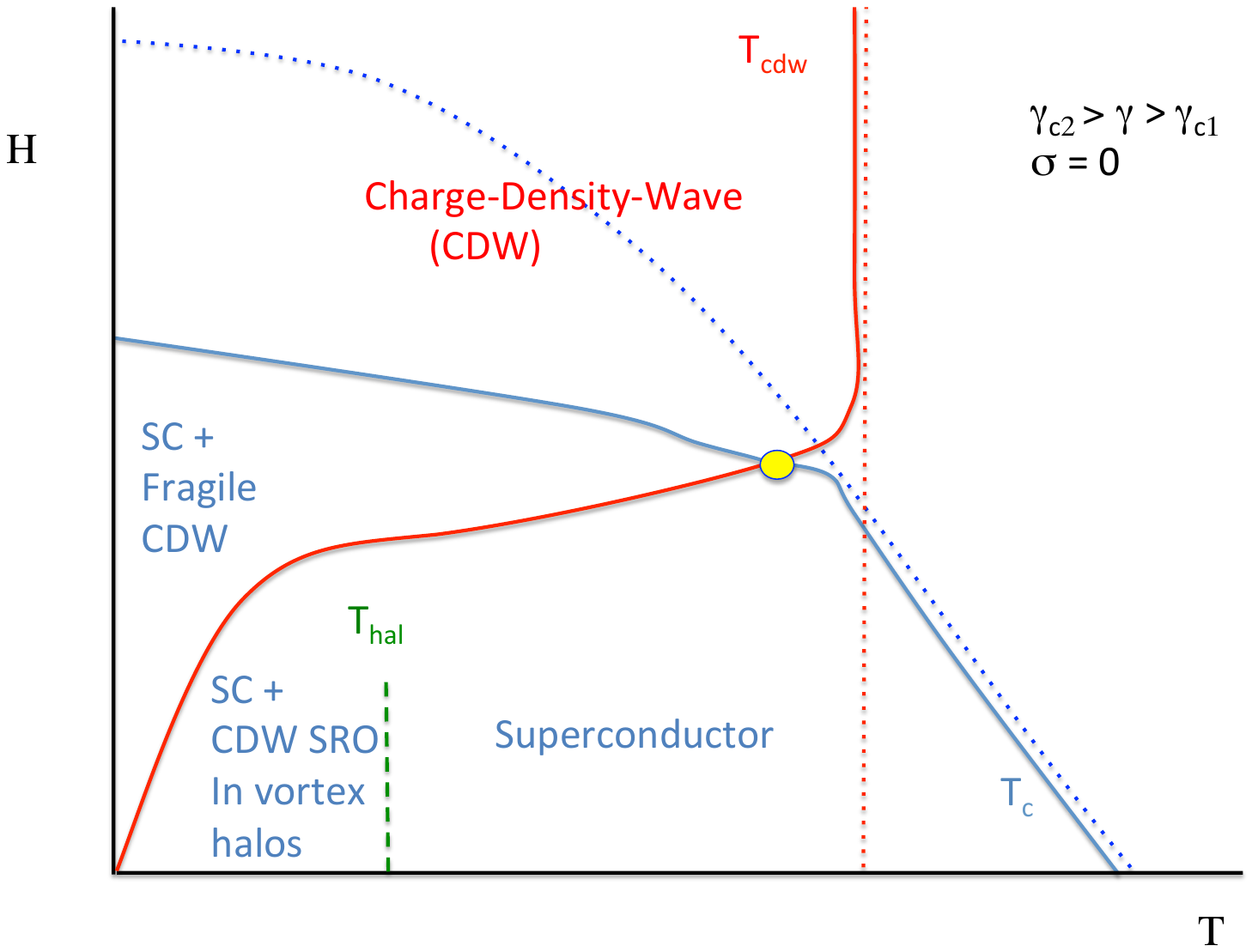}
\label{1a}}
\subfigure[]
{\includegraphics[scale=0.3]{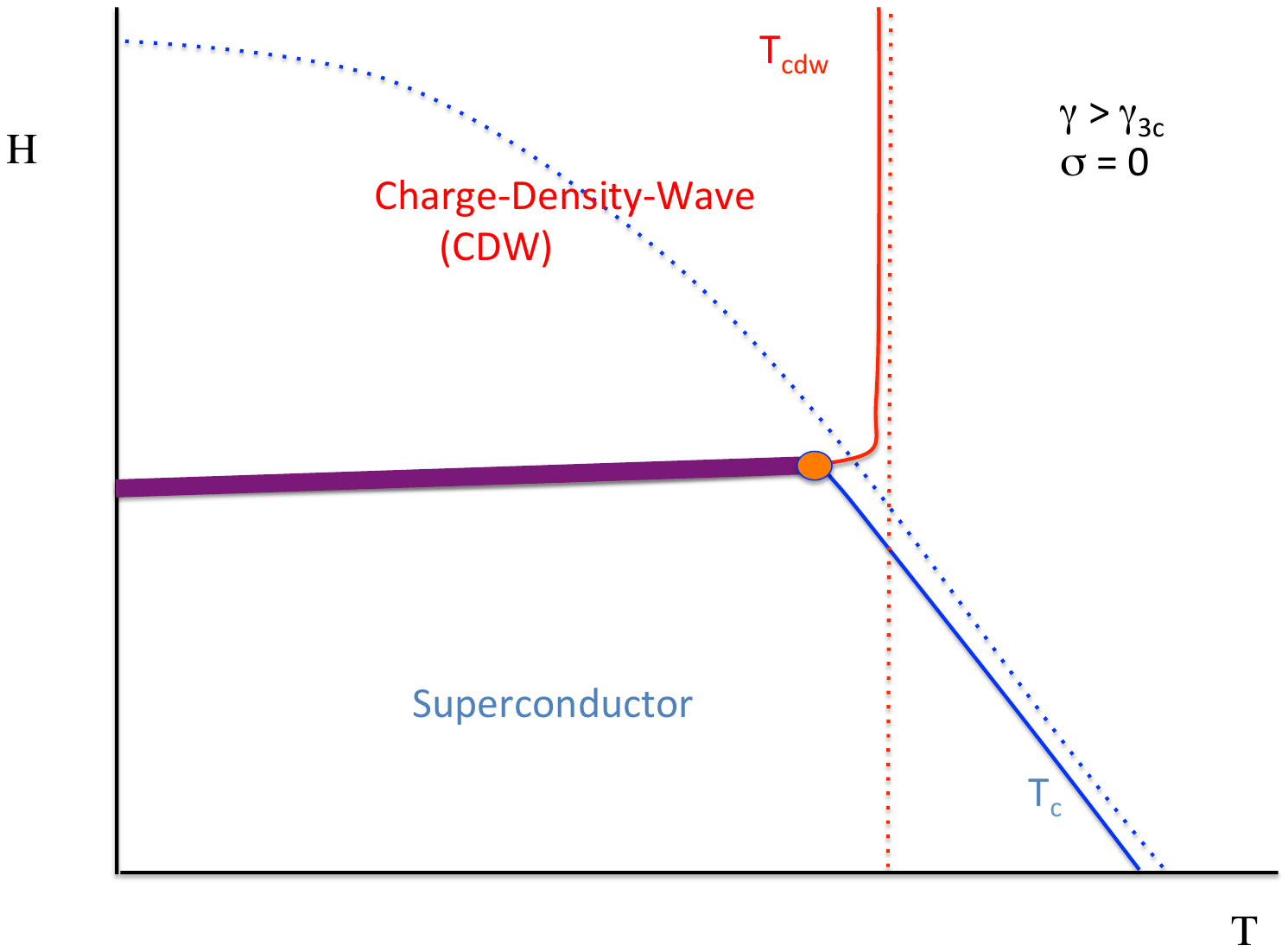}
\label{1b}}
\subfigure[]
{\includegraphics[scale=0.3]{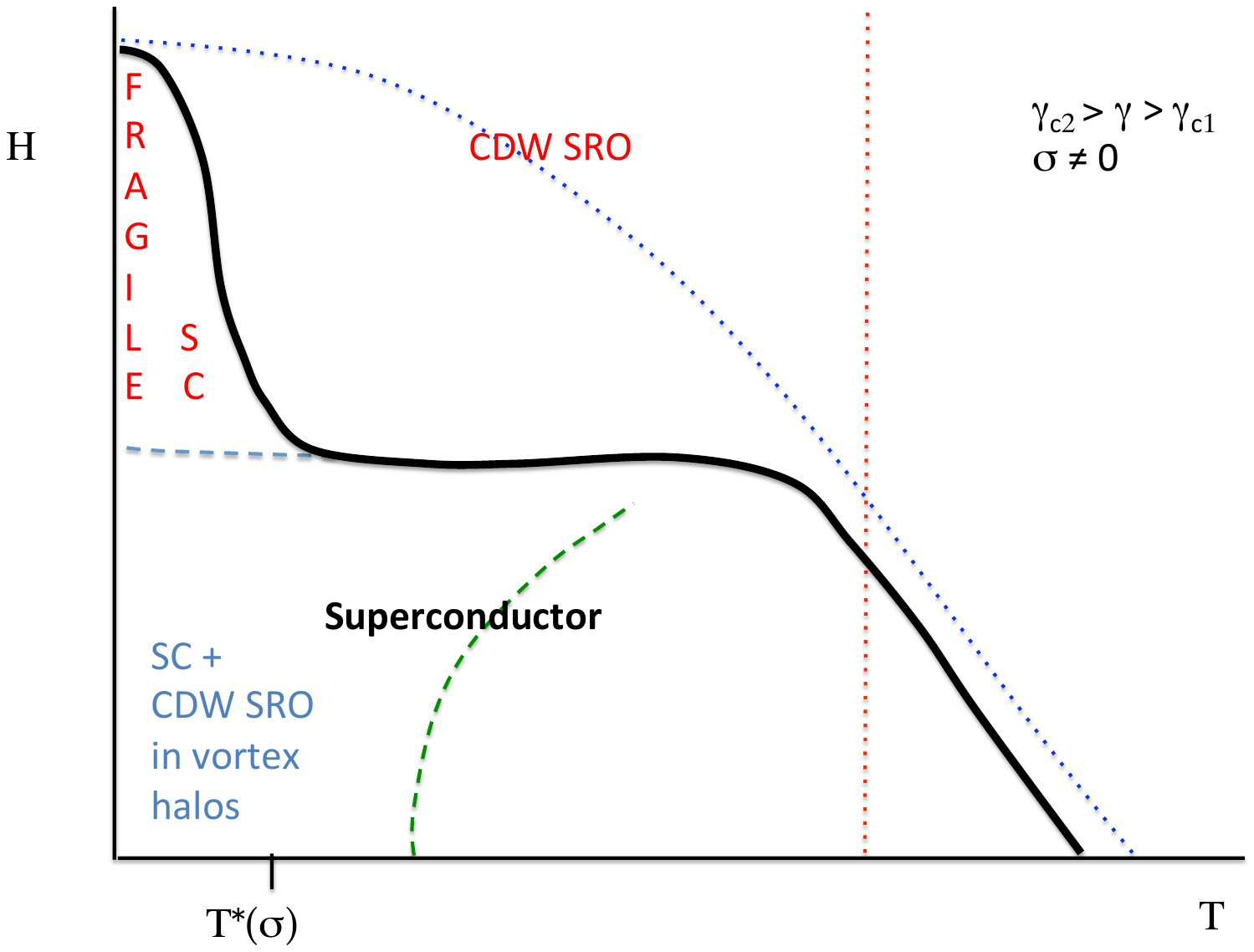}
\label{1c}}
\subfigure[]
{\includegraphics[scale=0.3]{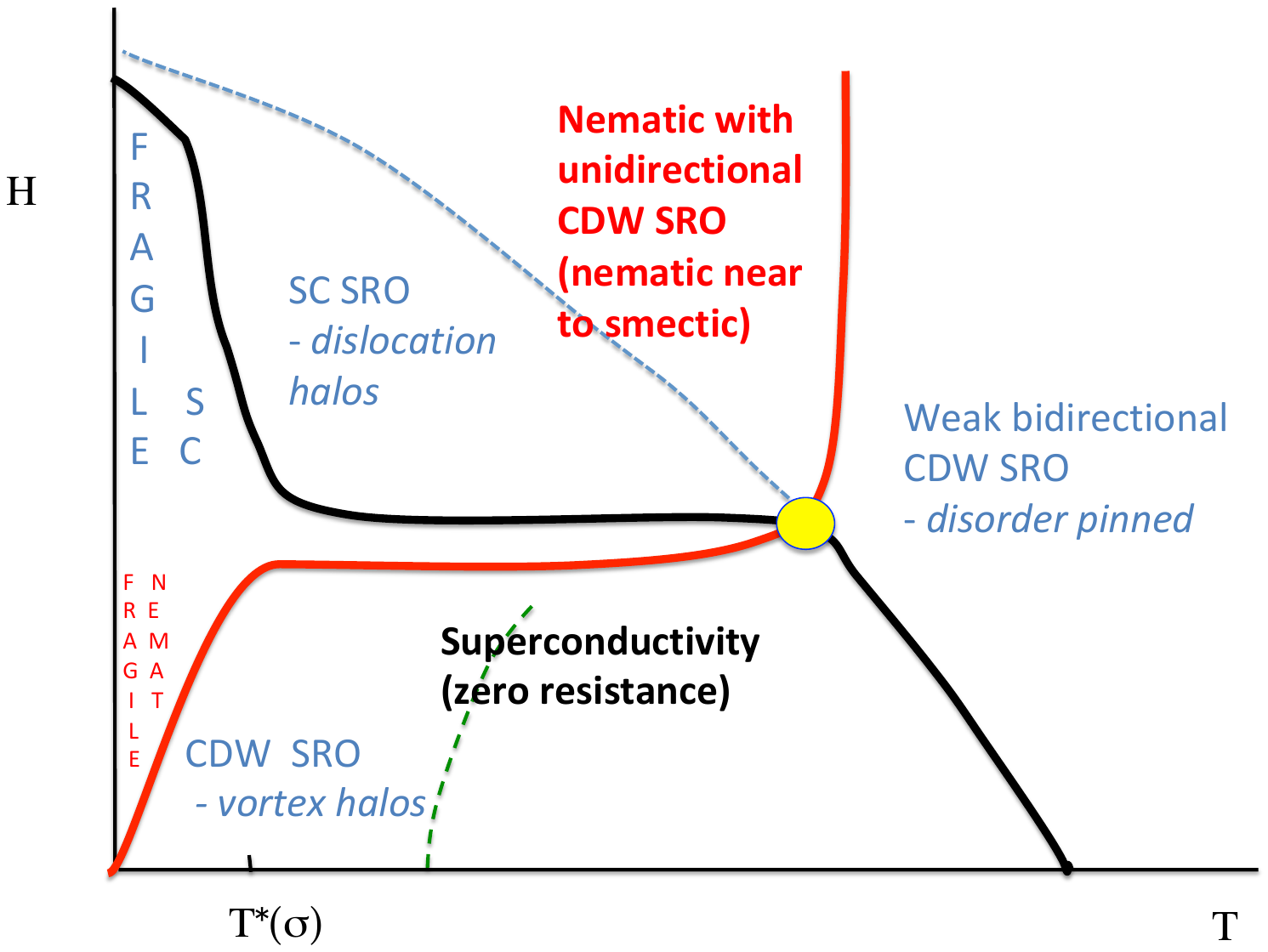}
\label{1d}}
\caption{ Schematic phase diagrams showing  competition between CDW and SC order in the absence of disorder (a and b) and in the presence of weak disorder (c and d).  
For reference, the dotted lines represent what would be the phase boundaries in the absence of coupling between the SC and CDW orders ($\gamma=0$).  The thin solid lines represent continuous and the heavy purple line  discontinuous transitions.  The  dashed lines indicate crossovers.  For further explanation 
see Sec. \ref{Qualitative}.}
\label{fig:phase}
\end{figure*}

Many qualitative aspects of our results are represented in the schematic phase diagrams in Fig. \ref{fig:phase}, which show a variety of ways the competition between CDW and SC order plays out;  these aspects can be motivated independently of the specific method of solution.  These figures are derived from the effective field theory defined below in the sense that we have shown that phase diagrams with these precise topologies and general shapes occur for appropriate choices of  parameters.   However, given that the parameters that enter an effective Hamiltonian are themselves generally functions of $T$ and $B$, albeit smooth, analytic functions, details of the global shapes of phase boundaries derived from such calculations are inessential.  Moreover,  the various features of the phase diagram are best computed  using different theoretical methods.  We have therefore drawn the phase diagrams to best illustrate the points of physics rather than to report results of a single calculation.  So as to avoid subtleties that are special to strictly 2D systems, 
 it is assumed that the system in question is a layered, quasi-2D system with weak uniform couplings between planes.  Thus, the CDW and zero field SC phases
exhibit true  LRO 
and the vortices induced by a finite $B$ form an Abrikosov  lattice or 
(in the presence of disorder) a vortex glass up to non-zero temperatures.\cite{PhysRevB.43.130,RevModPhys.66.1125}  

In the following, $\gamma$ represents the strength of the repulsion between the CDW and SC orders and $\sigma$ is a measure of the strength of the disorder.  Explicit definitions are given when we define the effective field theory in Eqs. \ref{H} and \ref{sigma}, below.   
While at a microscopic level, the electronic structure is changed in a way that affects the tendency toward both SC and CDW order (especially in the case of ``unconventional'' SC), at the order parameter level, gauge invariance precludes any linear (``random field'') coupling between disorder and the SC order parameter, so in our model, disorder couples directly only to the CDW order parameter.  (It indirectly affects SC through the local competition with CDW order.)  In this section, we will consider explicitly only the case in which in the zero field limit, the SC ordering tendency is slightly stronger than the CDW.

As a function of increasing $\gamma$, there are a variety of qualitatively distinct regimes possible in the clean limit,  $\sigma =0$, some of which we will outline here.  As already mentioned,  $\gamma_{c1}$  is defined such that at  $H=0$,  CDW and SC order coexist at low enough $T$ so long as $\gamma_{c1}> \gamma$, while no such coexistence occurs for $\gamma \geq \gamma_{c1}$.  Under appropriate circumstances, there is a second critical value $\gamma_{c2} > \gamma_{c1}$ such that at $T=0$, a CDW halo forms about an isolated vortex (which is generated by an infinitesimal non-zero $H$) for $\gamma_{c2} > \gamma > \gamma_{c1}$ but not for $\gamma> \gamma_{c2}$.  We will also derive a third critical value $\gamma_{c3} > \gamma_{c2}$, which is the smallest value of $\gamma$ such  that for  any $\gamma > \gamma_{c3}$ there is no range of $T$ and $H$ in which CDW and SC  coexist.

\underline {$\gamma=\sigma=0$}: The dotted lines in all four panels of Fig. \ref{fig:phase} delineate the phase boundaries in the clean limit and in the absence of coupling between the order parameters.  In this limit, at temperatures below the red dotted line (which is vertical since $B$ does not couple significantly to CDW order), there is CDW LRO.  The blue dotted line is the boundary of the SC (Abrikosov lattice) phase.  The two phase boundaries cross at a decoupled tetracritical point, and there is a broad region of coexisting SC and CDW order.
The assumption that the two orders have comparable strength is reflected in the relatively small magnitude of $T_c-T_{cdw}$.

\underline {$\gamma> 0$ with $\sigma=0$}:  For small  $\gamma>0$, the coexistence regime persists, but its area is reduced.  In Fig. \ref{1a} we show a representative phase diagram for  $\sigma=0$ and 
$\gamma_{c2} > \gamma > \gamma_{c1}$ (where $\gamma_{c1}$ and $\gamma_{c2}$ were defined 
 above).  
 Because $\gamma>\gamma_{c1}$, the SC order prevents any CDW LRO at $H=0$;  however, because $\gamma < \gamma_{c2}$, at small but non-zero $H$, CDW halos form around each vortex below the crossover temperature, $T_{hal}$, shown as the vertical dashed green line in the figure.   CDW LRO in the coexistence phase occurs below the solid blue line;  it is highly inhomgeneous at low magnetic fields, triggered by the (relatively weak) coupling from one vortex halo to the next, and hence is labeled ``fragile.''  At higher fields,  the balance between CDW and SC is reversed, with the result that there arises a pure CDW phase with uniform order.  Note that at low temperatures,  the relatively low value of $H_{c2}$ is a consequence of competition with the CDW rather than a property of the SC state itself.

For large enough  $\gamma > \gamma_{c3} > \gamma_{c2}$, the coexistence phase is entirely quenched, as shown in 
 Fig. \ref{1b}.  
 The thick purple  line delineates a first order field driven SC  to CDW transition which ends in a bicritical point at which the phase boundaries of the pure SC (solid blue line)   and CDW phases (red solid line) meet.

In fact, there are many other possible topologies of this phase diagram  depending on parameters:  For instance, even for  $\gamma>\gamma_{c3}$, rather than a tetracritical point, one can have a first order line that extends to higher temperatures than either of the ordered phases.  For smaller values of $\gamma$, it is possible to have more than one multicritical point -- for instance, at the high temperature side of the phase diagram, the SC and CDW phase boundaries could meet at a bicritical point, but then at lower $T$ a region of two-phase coexistence could arise in one of several possible ways.  In the case in which $\gamma_{c1} > \gamma >0$, the phase boundary between the SC phase and the coexistence phase can hit the $H=0$ axis at a non-zero critical temperature, while for $\gamma_{c3} > \gamma > \gamma_{c2}$ (assuming that the initial tetracritical point remains stable) this same phase boundary will hit the $T=0$ axis at a non-zero critical value of the magnetic field.

\underline {
$\gamma_{c2} > \gamma > \gamma_{c1}$ and $\sigma_c \gg \sigma > 0$}:  Even when we entirely ignore the effects of pair-breaking, when the disorder strength exceeds a critical value, 
$\sigma_c$, the locally pinned CDW correlations are sufficiently strong that SC cannot arise - for $\sigma > \sigma_c$, no broken symmetry phases exist at any $T$.  

The solid black curve in Fig. \ref{1c} shows the only true phase boundary in the presence of weak but non-vanishing disorder, $\sigma_c \gg \sigma > 0$.  It is a SC to non-superconducting (``normal'') transition.  In any dimension $d \leq 4$, the presence of weak disorder precludes the existence of CDW LRO.  A Bragg-glass phase with CDW quasi LRO is in principle possible in $d=3$, but for a quasi 2d system, it seems unlikely that such a phase would occur robustly.  Similarly, no first order transition is allowed in $d\leq 2$, and correspondingly it is likely that even rather weak disorder eliminates the possibility of such a transition in a quasi 2d system.  Thus, this transition is  continuous.  At long distances, the SC phase is a ``vortex glass,'' a phase with non-vanishing long-range phase rigidity and  (to the extent that gauge-field fluctuations can be ignored) a superconducting version of an Edwards-Anderson order parameter.\cite{PhysRevB.43.130}

The upturn of the SC phase boundary at low $T$ is the most notable new aspect of this phase diagram.  This upturn is the promised consequence of the presence of topological defects.
 In the entire range of field and temperatures  between the blue dotted line and the solid black line, it is only the presence of strong, local CDW correlations that is quenching the SC order.  Thus, so long 
as the superconducting coherence length is not too long ($\kappa < \kappa_c$),  a locally superconducting halo forms about CDW dislocations where the CDW order vanishes.  Inevitably, a form of granular SC correlations is generated.  At elevated temperatures, the Josephson coupling between neighboring dislocation halos is small compared to $T$, so there are at best extremely weak macroscopic signatures of these SC correlations.  However, global SC phase coherence onsets below a disorder dependent characteristic temperature, $T^\star(\sigma)$, comparable to the Josephson coupling, $J$ between neighboring dislocations.  
 Note that the resulting SC phase is ``fragile,'' both in the sense that it is destroyed by thermal fluctuations at temperatures far smaller than the zero field $T_c$, but also in that the critical current of the SC state is set by $J(\xi_{cdw})$, and so is also increasingly small the weaker the disorder.

 \underline{Tetragonal crystal with $\gamma_{c2} > \gamma > \gamma_{c1}$ and $\sigma_c \gg \sigma > 0$}:
  In the interest of simplicity, all of the model calculations we perform in the present paper are for the case of an orthorhombic crystal, where the direction of the CDW order is uniquely determined.  However, here and in the conclusion, we include a brief discussion of how the phase diagram differs in 
 a tetragonal crystal where even if the preferred CDW order is unidirectional, it can condense in either of at least two symmetry related directions.
 \footnote{If the preferred CDW order is not along a symmetry direction, then there are four symmetry related directions the CDW order can chose.}
In this case, for $\sigma=0$, CDW LRO breaks not only translation symmetry, but also a discrete rotational (or mirror) symmetry of the  crystal.  For non-zero $\sigma$, the translation symmetry aspect of the CDW order is lost, but the discrete point-group symmetry breaking aspect  survives up to a critical disorder strength.  Thus, in this case, as illustrated in Fig. \ref{1d}, even for non-zero $\sigma$ there can be a well defined thermodynamic phase transition to a state with ``vestigial nematic order,'' that is to say in which there is long-range CDW orientational order without positional order.\cite{nie}  Consequently, what appear as CDW crossover lines in the orthorhombic case shown in Fig. \ref{1c}, become the red solid phase boundaries in Fig. \ref{1d} in the tetragonal case.   We will return to this case in the final section.

\section{The Model}
\label{Model}
To make the discussion explicit, and since the qualitative behavior we are exploring is relatively insensitive to microscopic details, we begin by considering the properties of a minimal classical Landau-Ginzburg-Wilson (LGW) effective field theory.  
We consider two complex scalar fields,  a charge $2e$ field $\Delta$ representing the local SC order parameter, and $\phi$ representing the amplitude of a unidirectional incommensurate CDW.  (Generalizing the considerations to bidirectional CDW order, or multicomponent SC orders is straightforward, although not entirely without qualitatively new features.)  
We will focus on a  quasi-2D limit in which there are only  weak couplings between neighboring 2D layers of a 3D crystal.
 
 The effective free energy density of a plane is 
\bea
{\cal H}_{2d} =&& \frac {\kappa}2 \left|\left(- i \vec \nabla - 2e \vec A\right)\Delta\right |^2 - \frac{\alpha_\Delta}2 \left| \Delta\right |^2 + \frac {|\Delta|^4} 4 \nonumber \\
&&+  \frac {\kappa_{x}}2 \left|\partial_x\phi\right |^2 + \frac {\kappa_{y}}2 \left|\partial_y\phi\right |^2- \frac{\alpha_\phi}2 \left| \phi\right |^2 + \frac {|\phi|^4} 4 \nonumber \\
&& + \frac {\gamma}2 \left| \Delta\right |^2\left| \phi\right |^2 + h^\star \phi + \phi^\star h.
\label{H}
\eea
The weak couplings between order parameters ($\phi_n$ and $\Delta_n$ in neighboring planes ($n$ and $n+1$), are taken to be
\be 
{\cal H}_\perp  =- \frac{J_\phi}{2} [\phi_{n+1}^\star\phi_n + {\rm H. C.}]- \frac{J_\Delta}{2} [\Delta_{n+1}^\star\Delta_n + {\rm H. C.}].
\ee
This is at best an effective Hamiltonian, obtained by integrating out microscopic degrees of freedom, which means that even the parameters that enter the model should properly be taken to have analytic dependences on temperature, $T$, and magnetic field, $\vec H = \vec \nabla \times \vec A$.  Moreover,  in general there should be higher order terms, both in powers of the fields and in powers of derivatives, and at low enough temperatures, the quantum dynamics of the fields can be important.   

Without loss of generality, we have chosen units of  the two order parameters 
to set the coefiscients of $|\phi|^4$, $|\Delta|^4$ 
  to 1.  Moreover, we can chose units of distance and energy so that $\alpha_{\Delta}=\kappa=1$.  Because $\phi$ is a unidirectional CDW, $\kappa_x$ and $\kappa_y$ need not be equal, even in a tetragonal system, but for simplicity we will set $\kappa_x=\kappa_y =1$.  In the above, $h$ is a gaussian random complex field representing the effect of quenched disorder, with
\be
\overline{ h(\vec r)}=0 \ \ \overline{ h(\vec r) h(\vec r^\prime)}= 2\sigma^2 \delta(\vec r-\vec r^\prime),
\label{sigma}
\ee
the overline represents the configuration average, and $\sigma$ characterizes the strength of the disorder.  Importantly, the magnetic field couples directly only to the superconducting order, while the disorder couples only to the CDW order.  
Since we will in particular focus primarily on the case in which superconductivity is the dominant zero field order,  but  in which CDW is only weakly subdominant, we will assume $1 \gg 1 - \alpha_{cdw} > 0$. We will also assume that the repulsion between the two orders is reasonably strong, $\gamma  \sim 1$.

Ideally, we would treat the 
\be
S = \beta \sum_n\int d\vec r \left[{\cal H}_{2d}+ {\cal H}_\perp\right]
\ee
 as the effective action where $e^{-S}$ gives the Boltzman weight for each field configuration.  Below we carry this out in several approximate ways.
 \footnote{There have, needless to say, been many earlier studies of competing orders in the context of just such a LGW effective field theory.  In particular, the problem in the absence of the gauge-field coupling ($\vec A=\vec 0$) and in the absence of disorder is the standard model for analysis of bi and tetra-critical points.  For the state of the art theory of this, with references to the earlier literature, see P. Calabrese, A. Pelissetto, and E. Vicari, \prb {\bf 67}, 054505 (2003).}

\section{Saddle point solution -- Landau-Ginzburg theory}
\label{meanfield}
The field configurations that minimize $S$ determine the classical ground-state ($T=0$) configurations of the fields.  (More generally, if we treat $S$ as  the Landau-Ginzburg (LG) free energy, in which the coefiscients $\alpha_\Delta$ and $\alpha_\phi$ and possibly others are taken to be $T$ dependent, minimizing $S$ corresponds to LG mean field theory.)

\subsubsection{$\sigma=0$}
In the following discussion, we focus on $\gamma_{c1}<\gamma< \gamma_{c2}$. For $H=0$, the ground-state has SC order and vanishing CDW order. In this case, the CDW order develops, and increases in strength with increasing magnetic fields. However, the magnetic field produces a vortex crystal  and suppresses the overall SC order - most strongly in vortex cores.  Needless to say, the CDW order develops a spatially varying magnitude in response to the field-induced vortex lattice.  No analytic solution of this modulated state exists.

However, the state is increasingly homogeneous as $H$ approaches the upper critical field, $H_{c2}$, defined such that 
 SC order is quenched for $H> H_{c2}$:
\be
H_{c2} 
= H_{c2}^{(0)}[1 - \gamma \alpha_\phi/\alpha_\Delta]
\label{Hc20}
\ee
where $H_{c2}^{(0)} = (2e)^{-1}\alpha_\Delta$ is the value of $H_{c2}$ in the absence of competition with the CDW.

\subsubsection{$H_{c2}^{(0)}> H> H_{c2}$, and $\sigma>0$}\label{dislocation}
In the presence of weak disorder, dilute pinned dislocations disrupt  the CDW order, 
opening the possibility of local superconductivity. 
 We now consider the problem of superconductivity 
 at an isolated CDW dislocation. If there is a non-trivial solution for this problem, long-range superconductivity will always  develop  at low enough temperature in the presence of infinitesimal disorder  up to a critical  magnetic field $ H_{c2}(\sigma) > H_{c2}\equiv H_{c2}(\sigma=0)$. 
 
\begin{figure}[htb]
\centering
\includegraphics[width=8cm]{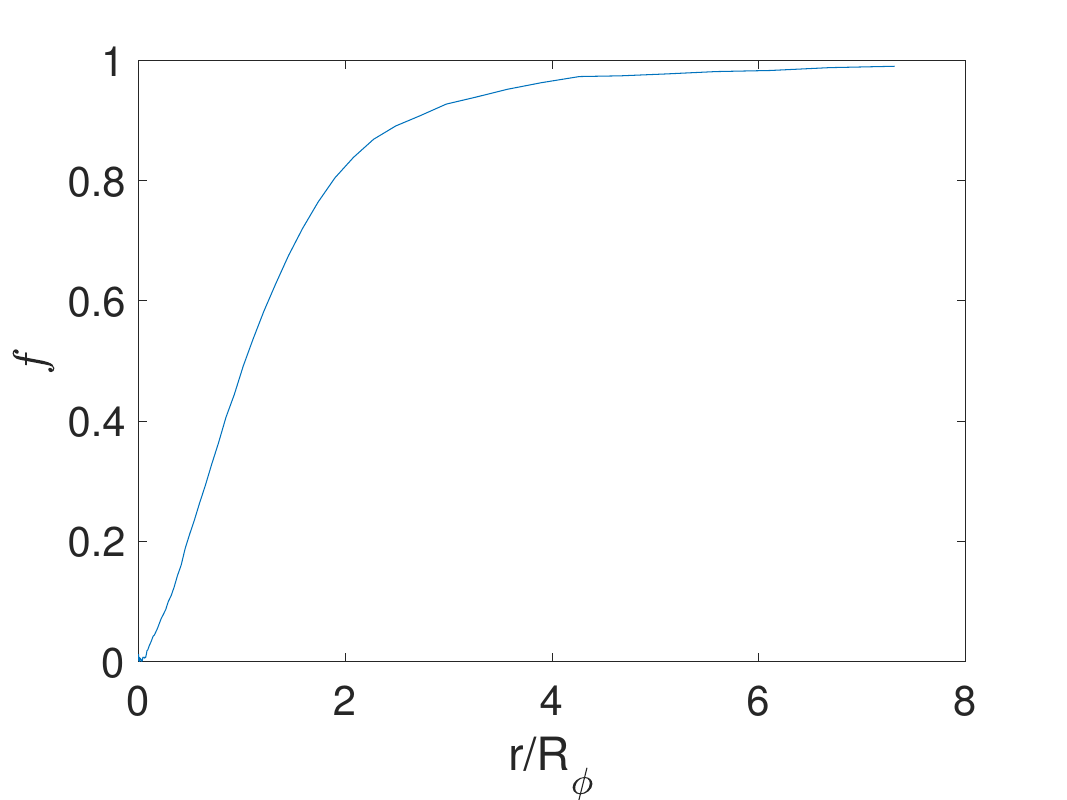}
\caption{The magnitude of CDW order inside a single  
 dislocation as a function of the distance from the center obtained from numerically minizing ${\cal H}$ in  Eq. \ref{e3}. The characteristic length is $R_{\phi}=\sqrt{
 {\kappa_\phi}/{\alpha_\phi}}$.  
  }
\label{f1} 
\end{figure}

Around the critical magnetic field, $\Delta$ is 
small, so the CDW profile can be firstly solved while neglecting superconductivity. Superconductivity can then be studied  treating the CDW as a fixed potential. Neglecting all $\Delta$ terms, the simplified Hamiltonian for the CDW is
\begin{equation}
{\cal H}[\phi]=\frac{\kappa_\phi}{2}|\bm{\nabla}\phi|^2-\frac{\alpha_\phi}{2}|\phi|^2+\frac{1}{4}|\phi|^4.
\label{e3}
\end{equation}
We assume an ansatz for an isolated dislocation, with the following boundary condition
\begin{equation}
\begin{split}
&\phi(r,\theta)=\sqrt{
\alpha_{\phi}} \ f(r)e^{i\theta}\\
&f(r\rightarrow{\infty})=1\\
&f(r\rightarrow{0})=0.
\end{split}
\end{equation}
The solution $f(r)$ can be obtained numerically, as shown in Fig.\ref{f1}. The characteristic length of the dislocation is $R_{\phi}=\sqrt{
 {\kappa_\phi} / {\alpha_\phi}}$. 

Near the critical field, where SC is weak, the back reaction of the SC correlations on the form of the dislocation can be ignored.  Thus, the saddle point equation for $\Delta$ is
\bea
&&\hat{O} \Delta=-
({1}/{2})\ |\Delta|^2{\Delta} \\
&&\hat{O} \equiv\left[-\kappa_\Delta(-i\bm{\nabla}-2e\bm{A})^2-\alpha_\Delta+(\gamma\alpha_\phi){f^2}\right].
 \nonumber
\eea
In order to have a nontrivial solution, the operator $\hat{O}$ 
 must have at least one negative eigenvalue. This problem 
 is equivalent to the Schrodinger Eq. for a charged particle in a magnetic field $H$ and a potential $V(r)=\gamma\alpha_\phi{|f(r)|^2}$. 

\begin{figure}[htb]
\centering
\includegraphics[width=8cm]{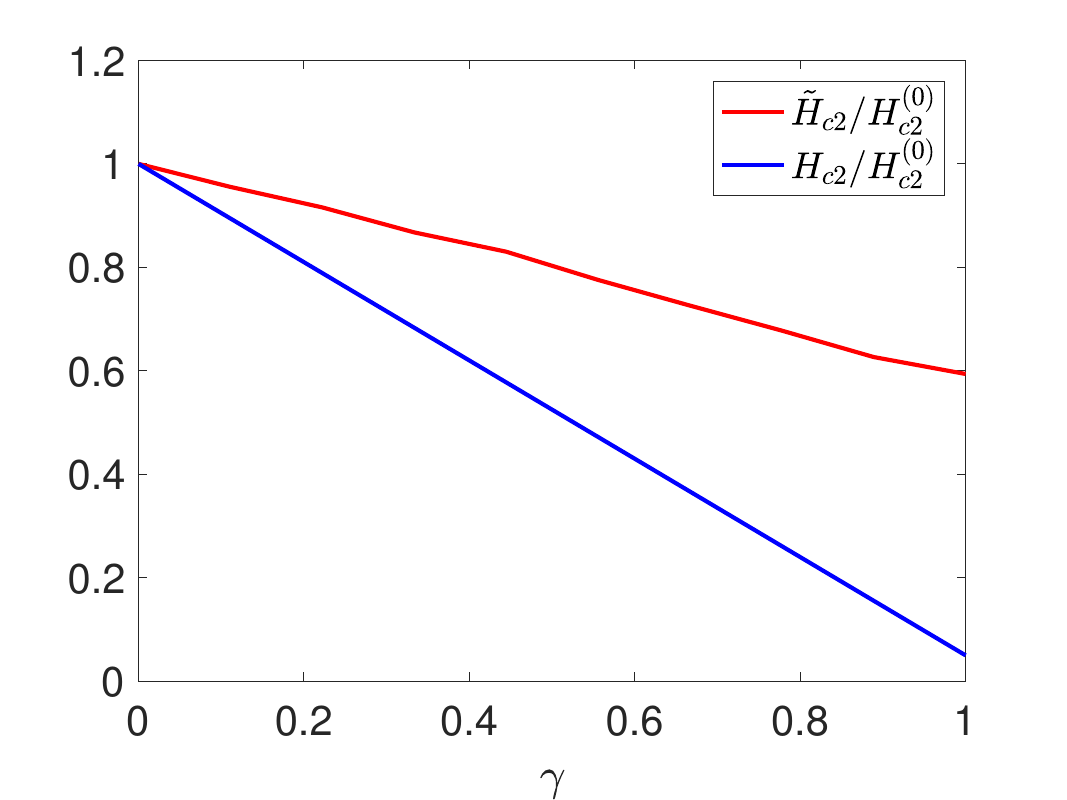}
\caption{Upper critical fields $H_{c2}$ (in the absence of disorder) and $\tilde H_{c2}$ (in the presence of dilute, disorder induced dislocations) as a function of $\gamma$ as computed from solution of the Landau-Ginzburg equations.  Here $H_{c2}^{(0)}$ is the value of the critical field in the absence of competition with CDW order ($\gamma=0$).  Here we have taken $\alpha_\phi=0.95\ \alpha_\Delta$ and $\kappa_\Delta=\kappa_\phi=1$.  
}
\label{fh} 
\end{figure}

If we work in symmetric gauge, the solutions can be classified by their out-of-plane angular momentum, $m$, with the lowest energy solution lying in the $m=0$ sector.  In this sector
\begin{equation}
\hat{O}_{m=0}=-\kappa_\Delta\bm{\nabla}^2-\alpha_\Delta+\gamma\alpha_\phi{f^2}+\kappa_\Delta{e^2B^2}r^2
\label{eqo}
\end{equation}
 The critical field $\tilde H_{c2}$ can be obtained numerically as the point at which the lowest eigenvalue vanishes. Representative results are shown in Fig.\ref{fh}; as expected,   
 $ \lim_{\sigma\to 0} H_{c2}(\sigma) = \tilde H_{c2} >H_{c2}$. 

The fact that for non-zero $\gamma$, $\tilde H_{c2} > H_{c2}$   implies a strikingly  non-analyticity behavior of $H_{c2}(\sigma)$.  Specifically, since we expect a non-zero concentration of dislocations for any non-zero value of $\sigma$,  
\be
\lim_{\sigma\to 0} H_{c2}(\sigma) = \tilde H_{c2} > H_{c2}(\sigma=0) \equiv  H_{c2}.
\label{Hsigma}
\ee

With increasing $\sigma$, the concentration of dislocations increases, and consequently one expects a range  in which $H_{c2}(\sigma)$ is an increasing function of $\sigma$ as the halos begin to overlap.  However, as we will see below, for large enough $\sigma$, there is no superconductivity even at $H=0$, which implies the existence of a critical disorder strength, $\sigma_c$ such that $H_{c2}(\sigma) \to 0$ as $\sigma \to \sigma_c$ from below. (See Fig. \ref{fh2})

\section{Variational solution }
\label{fluct}
A first step beyond LG theory is obtained by treating fluctuations approximately using the Feynman  approach, in which we introduce a quadratic trial Hamiltonian, with parameters optimized according to a variational principle.  To treat the disorder averages properly, we introduce $n$ replicas of the theory, evaluate the trace over disorder fields $h$ as if they were in equilibrium, and then take the limit as $n\to 0$.  (Details of this procedure are reported in Ref. \onlinecite{nie}.) 
The replicated trial Hamiltonian density in the normal phase (i.e. in the absence of broken symmetry) is then of the form
\begin{equation}
\begin{split}
&{\cal H}_{tr}=\sum_a^n\Big[\frac{\kappa_\phi}{2}|\bm{\nabla}\phi_a|^2+\frac{\mu_\phi}{2}
|{\phi}_{a}|^2\\
&+\frac{\kappa_\Delta}{2}|(-i\bm{\nabla}-2e\bm{A})\Delta_a|^2+\frac{\mu_\Delta}{2}
|\Delta_{a}|^2\Big]\\
&-\beta\sigma^2\sum_{ab}^n
{\phi}_{a}^\star
{\phi}_{b}.
\end{split}
\end{equation}
where, after taking the $n\to 0$ limit, the self-consistency for the variational parameters, $\mu_\Delta$ and $\mu_\phi$, 
become
\begin{equation}
\begin{split}
&\mu_\phi=-\alpha_\phi+\frac
{(N+2)}{N}
\langle
|{\phi}_{1}|^2\rangle+\gamma\langle
|{\Delta}_{1}
|^2\rangle\\
&\mu_\Delta=-\alpha_\Delta+\frac{(N+2)}{N}
\langle
|{\Delta}_{1}|^2\rangle+\gamma\langle
|{\phi}_{1}|^2\rangle
\end{split}
\label{e18}
\end{equation}
and where $N=2$.  (We have introduced $N$ to signify the number of components of the order parameter fields, as there is an interesting large $N$ limit in which the variational approach becomes exact, but we will always work with $N=2$ in the following.)  These equations are valid so long as no symmetries are broken.  The expectation values that enter these equations are taken with respect to the trial Hamiltonian, and so depend (in a complicated non-linear manner) on the variational parameters.

The mean-square density wave fluctuations are the sum of two terms - the first reflecting  the disorder induced fluctuations and the latter the thermal fluctuations:
\begin{equation}
\begin{split}
&\langle
|{\phi}_{1}|^2\rangle=4
\sigma^2\int\frac{d^3k}{(2\pi)^3}{\frac{1}{(\mu_\phi+\kappa_\phi{k_\parallel^2}+2J_\phi\cos{k_\perp})^2}}\\
&+
2T\int\frac{d^3k}{(2\pi)^3}{\frac{1}{\mu_\phi+\kappa_\phi{k_\parallel^2}+2J_\phi\cos{k_\perp}}}
\end{split}
\label{phi2}
\end{equation}
In the absence of disorder, the CDW transition temperature, $T_{cdw}$, is the point at which $\mu_\phi \to 2J_\phi$.  However, notice that the disorder induced fluctuations diverge as $\mu_\phi \to 2J_\phi$, which (correctly) encodes the fact that for $\sigma\neq 0$, CDW order is precluded in $d\leq 4$.
The mean square superconducting fluctuations do not depend explicitly on the disorder, but are different in the presence and absence of a magnetic field.  For $B=0$,
\begin{equation}
\begin{split}
&\langle
|{\Delta}_{1}|^2\rangle=2T\int\frac{d^3k}{(2\pi)^3}{\frac{1}{\mu_\Delta+\kappa_\Delta{k_\parallel^2}+2J_\Delta\cos{k_\perp}}}
\end{split}
\label{delta2a}
\end{equation}
while in the presence of a magnetic field, it can be expressed as a sum over Landau levels
\begin{equation}
\begin{split}
&\langle
|{\Delta}_{1}|^2\rangle=
\\
&\frac {2TBe}{\pi}\sum_{p=0}^\infty\int\frac{dk_\perp}{(2\pi)}{\frac{1}{\mu_\Delta+4eB\kappa_\Delta(p+1/2)+2J_\Delta\cos{k_\perp}}}
\nonumber
\end{split}
\label{delta2b}
\end{equation}
The value of $T_c$ is extracted from the self-consistency equations as the point at which $\mu_\Delta \to 2J_\Delta -2e\kappa_\Delta B$.

The variational approach can be extended to the ordered phase by adding an explicit symmetry breaking field. For instance, for the case of an ordered CDW (in the absence of disorder), a term of the form $\frac{m}{2}[\phi+\phi^\star]$ must be 
added to  ${\cal H}_{tr}$.  
More about applying this more general approach to the present problem is presented in the Appendix.

\subsection{Critical value of $\sigma$ at $T=0$}
\begin{figure}[htb]
\centering
\includegraphics[width=8cm]{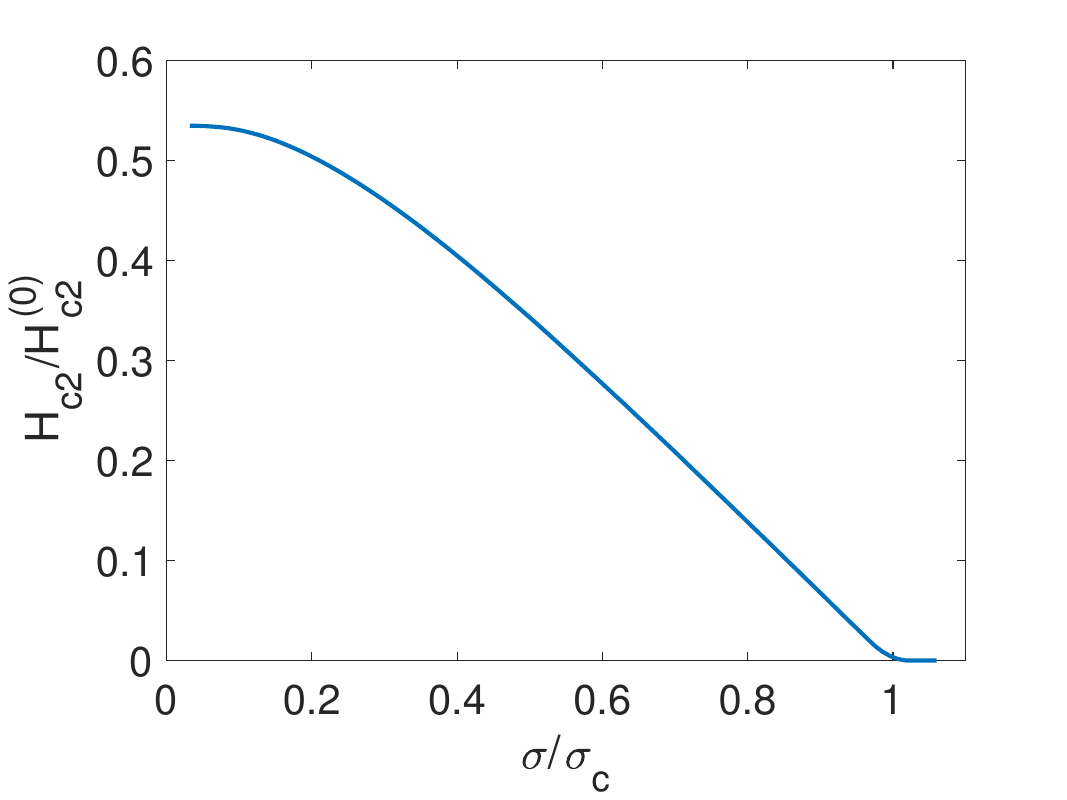}
\caption{
Variational result for $H_{c2}$ 
 as a function of $\sigma$.   
 Here $H_{c2}^{(0)}$ is the value of the critical field in the absence of competition with CDW order   
 ($\gamma=0$), $\sigma_c$ is the zero field critical disorder from Eq. \ref{sigmac}, and 
 we have set $\alpha_\phi=0.95\ \alpha_\Delta$, $\gamma=\alpha_\Delta=1$ and $\kappa_\Delta=\kappa_\phi=1$.  
}
\label{fh2} 
\end{figure}

As already mentioned, non-zero disorder always precludes CDW LRO.  From Eq. \ref{e18} it follows that  $\mu_\Delta$ decreases monotonically as a function of increasing disorder for fixed $T$ and $H$.  Thus in the variational solution, CDW disorder is always harmful for superconductivity. 
(This is not unexpected, as the replica trick does not explicitly incorporate the disorder-generated dislocations  that lead to the 
non-analytic enhancement of superconductivity by disorder at low $T$ and moderate $H$.)
The dependence of the $T=0$ critical field as a function of disorder, $H_{c2}(\sigma)$, that results from the present analysis is 
 shown in Fig. \ref{fh2}. Note that there is a critical disorder strength, $\sigma_c$,  where even though long-range CDW correlations have been entirely destroyed, the short-range amplitude of the pinned CDW order is sufficiently strong that it quenches superconductivity, even at $T=0$ and $H=0$. 

Specifically, at $T=0$, (no thermal fluctuations)
the self-consistent equation 
simplify to
\begin{equation}
\begin{split}
&\mu_\phi=-\alpha_\phi+2\langle|{\phi}_{1}|^2\rangle\\
&\mu_\Delta=-\alpha_\Delta+\gamma\langle|{\phi}_{1}|^2\rangle\\
&\langle|{\phi}_{1}|^2\rangle=4
\sigma^2\int\frac{d^3k}{(2\pi)^3}{\frac{1}{(\mu_\phi+\kappa_\phi{k_\parallel^2}+2J_\phi\cos{k_\perp})^2}}.
\end{split}
\end{equation}
 The solution for 
 these equations are
\begin{equation}
\begin{split}
&\mu_\phi=-\alpha_\phi+8\sigma^2\int\frac{d^3k}{(2\pi)^3}{\frac{1}{(\mu_\phi+\kappa_\phi{k_\parallel^2}+2J_\phi\cos{k_\perp})^2}}\\
&\mu_\Delta=-\alpha_\Delta+\frac{\gamma}{2}(\mu_\phi+\alpha_\phi)
\end{split}
\end{equation} 

The critical disorder strength $\sigma_c$ can be extracted from $\mu_\Delta \to \mu_{\Delta ,c} \equiv 2J_\Delta-2e\kappa_\Delta H $, i.e.
\begin{equation}
\begin{split}
&\sigma_c^2=\frac{\pi}{2}\kappa_\phi(\alpha_\phi+\mu_{\phi,c})\sqrt{\mu_{\phi,c}^2-
\mu_{\Delta,c}^2}\\
&\mu_{\phi,c}=\frac{2\alpha_\Delta+
\mu_{\Delta,c}}{\gamma}-\alpha_\phi.
\end{split}\label{sigmac}
\end{equation}
At $H=0$, the value of $\sigma_c$ extracted in this way agrees well with the results from the Monte Carlo calculations in the next section.

\subsection{$H_{c2}$ in the absence of disorder}
\label{hc2cdw}
The phase boundaries in the limit $\sigma=0$ are somewhat more complicated to derive.  In particular, the $H_{c2}$ line over much of the range of $T$ separates a non SC phase with CDW LRO from a phase with coexisting SC and CDW LRO.  However, while for $H< H_{c2}$, the coexisting phase is a spatially inhomogeneous Abrikosov lattice, the $H_{c2}$ line itself can be approached from above, where the system remains spatially uniform.  Thus the only new complication is that  this involves treating the problem in the presence of a broken symmetry, as discussed in the Appendix.  Representative results of this analysis are  shown as in Fig. \ref{f2}.

The nature of the multicritical points that occur in this limit is still more subtle.  Often even when the mean-field phase diagram suggests a single tetra-critical point, the variational approach yields weakly first order transitions and a still more complex phase diagram.  This is likely unphysical, and in any case affects only a very limited range around the multicritical point.  For this reason, in presenting the results of the variational calculation in Fig. \ref{f2}, we have ignored these subtleties, and have instead presented only the phase boundaries derived under the assumption that all transitions are continuous.  For reasons discussed in the Appendix, these results are computed in the presence of non-vanishing interplane couplings,  $J_\Delta=J_\phi=0.3$. 

\section{
Classical Monte Carlo Results}
\label{mc}
\begin{figure}[htb]
\centering
\includegraphics[width=6cm]{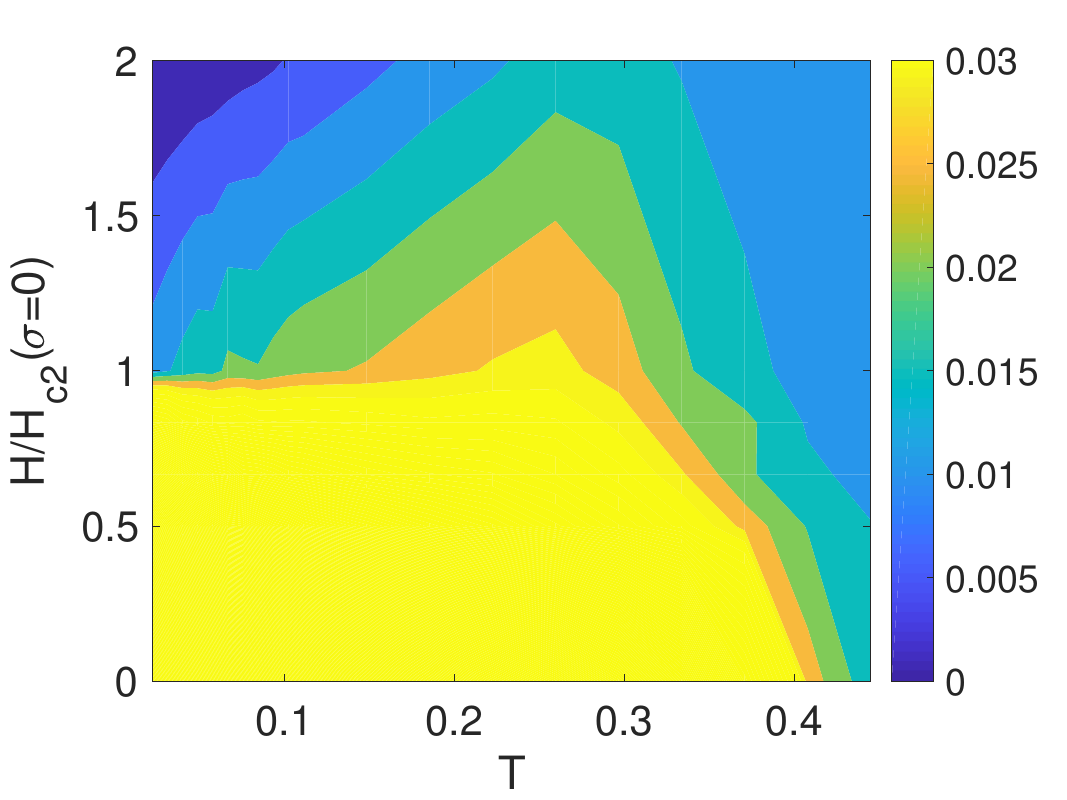}
\includegraphics[width=6cm]{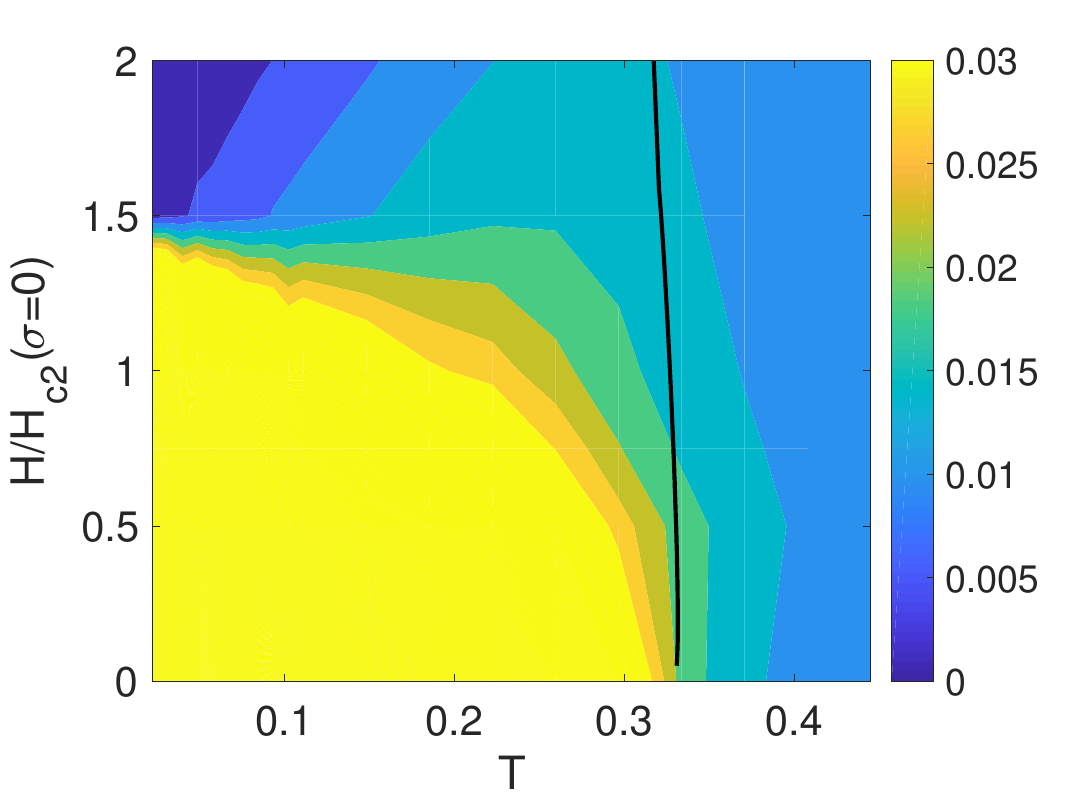}
\caption{
Monte Carlo results for the magnitude of the 
SC pair-field correlator, $G_{SC}$, at large distances as defined in Eq. \ref{emc}. Parameters are given in the first paragraph of Sec. \ref{mc}. The bright region (where this correlation is large) 
roughly corresponds to the ordered SC phase, while the dark region is the disordered phase. $H_{c2}(\sigma=0)=H_{c2}$ is the upper critical magnetic field in the absence of disorder, obtained in Eq. \ref{Hc20} and \ref{Hsigma}. 
Top panel:  results with zero disorder, $\sigma=0$.  
Bottom panel: results for weak disorder, $\sigma=0.05\sigma_c$.  The solid line is the SC phase boundary obtained from the variational calculation for $J_\Delta=J_\phi=0.3$. The summation over $k_\parallel$ in Eq. \ref{phi2} is capped at $k_{\parallel,\text{max}}=20\pi$, and the summation over $p$ in Eq. \ref{delta2a} is capped at $4eB(p_{\text{max}}+\frac{1}{2})=(k_{\parallel,\text{max}})^2$.
Notice that, consistent with theoretical expectations, weak disorder  decreases the range of temperature in which strong superconducting correlations survive (i.e. decreases the effective  critical temperature) but increases the extent of superconducting correlations at low fields (i.e. increases the effective the critical magnetic field). 
 While the variational calculation 
 captures some aspects of the exact results at low fields and $T$ near $T_c$, it manifestly 
  fails to explain the behavior at higher fields where CDW dislocations and possibly also vortex fluctuations play an essential role in the physics. }
\label{f2} 
\end{figure}

Finally, we treat $e^{-S}$ as the Boltzman weight and compute the phase diagram that results by exact numerical Monte-Carlo evaluation of thermodynamic correlation functions.  Of course, the down side of this is that it does not give analytic insight, and requires specific choices of model parameters, but it does allow us to verify the qualitative validity of 
some of the inferences made on the basis of the approximate analytic results discussed above.
In the following 
we set $\kappa_\Delta=\kappa_\phi$, 
$\alpha_\phi=0.95\ \alpha_\Delta$, and $\gamma = \alpha_\Delta$. Since the  calculations are carried out in 2D, this corresponds to setting the interplane couplings, $J_\Delta \ {\rm and} \ J_\phi=0$.

To permit numerical solution, we discretize the continuum Hamiltonian, but with  lattice constant (which is not physically meaningful)  chosen smaller than the coherence lengths. The simulation is set on a square lattice with $(L_x,L_y)=(20,100)$ and periodic boundary conditions. The vector potential $\bm{A}$ is set to be $(-By,0)$. The Monte Carlo simulation is performed for 400000 steps, and measurement is done on the last 320000 steps. For systems with disorder, the whole process is averaged over 60 different disorder configurations.  

Because we are studying a 2D model, for any non-zero $T$ there can be no true long-range-order, and for non-zero $H$ and $\sigma$, there should be no sharply defined superconducting or CDW phase.  As a suitable proxy to compare with the analytic theories, we thus defint  the following quantities:
\begin{equation}
G_{X}(\vec R)\equiv\overline{\frac{1}{L_xL_y}\sum_{i}|\langle{X_iX_{i'}}\rangle|},
\label{emc}
\end{equation}
where $X_i$ is the SC or CDW order parameter on each site $i$. $i'$ is chosen for each $i$, such that the 
displacement between them is $\vec R$, $\langle \ \rangle$ represents the thermal average and the overline represents the average over disorder configurations. Note that for 
each value of the magnetic field, 
the lattice constant is 
chosen to allow periodic boundary conditions. We choose $|\vec R|$ to be the largest possible distance between two points, in the system with  smallest lattice constant (i.e. largest magnetic field). We first perform thermal averaging to obtain the correlation function over sites $i$ and $i'$. Then we take average of the absolute value of the correlation function for all the pairs of sites.  This is a physically meaningful measure of the strength of the CDW and SC correlations, and relatively less sensitive to the presence or absence of small interplane couplings.

We now compare the phase diagram for zero disorder and $\sigma=0.05\sigma_c$, where $\sigma_c$ is defined in Eq. \ref{sigmac} for H=0. Here we plot the order parameter $\langle{\Delta}\rangle$ as a function of temperature and magnetic field, as shown in Fig.\ref{f2}. 
As expected, the disorder suppresses the superconducting $T_c$, but at low $T$ enhances the range of magnetic field over which SC survives, thus confirming the most dramatic qualitative expectation from the above theory.  On the other hand, the mean-field phase boundary shown as the solid line in the lower panel of the figure clearly overestimates the strength of the superconductivity correlation at large magnetic fields.  Conversely, since our diagnostic of SC correlations in the Monte Carlo results is crude, the full extent of the fragile superconducting phase at low $T$ and high $H$ (which by its very character has rather small amplitude mean superconducting correlations) is not visible in Fig. 5.

\section{Concerning the cuprate phase diagram}
\label{htc}

In a complex material such as the cuprates, various microscopic and material specific features always complicate any theoretical analysis.  Even at the level of order parameter theories, there are more players than the SC and a single component CDW order.  To begin with, since the Cu-O plane is approximately tetragonal, it is necessary to include  at least two CDW orders, with ordering vectors related by a $C_4$ rotation.  In various parts of the phase diagram - as well as in serious studies of model problems\cite{zaanen89,schulz90,machida-1989,white-1998,hellberg-1999,hellberg-1997,RevModPhys.75.1201,vojta} such as the Hubbard model which are thought to capture some of the essence of the microscopic physics of the cuprates - there is also clear evidence of incommensurate SDW order, and suggestive evidence of PDW order.\cite{himeda,rice,poilblanc,corboz,PhysRevB.97.174510}  
An additional complication is that  at $T\to 0$, the presence of gapless quasiparticles leads to non-local interactions and of course quantum fluctuations of the various order parameters need to be included;  none of these features are captured in the LGW framework adopted above.

Still, in certain cuprates, there is a range of doping in which the only two orders that have been clearly identified are CDW and SC order.\cite{julien-pnas,haug-2010,giacomo-2012,chang-2012,huecker-2014,comin-2015,gerber-2015,chang-2016,jang-2018}  It is thus interesting to explore to what extent the salient features of the phase diagram in this regime can be understood, following the above considerations, as an expression of the competition between these two forms of order.  
To get closer to the physics of the cuprates, 
we generalize the above discussion to the case of the competition between CDW and SC orders in a tetragonal system, where there are two CDW order parameters, $\phi_x$ and $\phi_y$, which are the slowly varying CDW amplitudes for ordering vectors along the x and y axes, respectively.  Following the analysis in Refs. \cite{nie,akash}, we assume that there is a strong repulsive interaction between these two components of the CDW order, so that the preferred ordered state is undirectional (striped) rather than bidirectional (checkerboard).   
When we generalize the above model to include both $\phi_x$ and $\phi_y$, the bulk of the considerations go through in analogous fashion. As already mentioned in the introduction, the most important difference is that in addition to breaking translational symmetry (in one direction), a stripe ordered CDW also spontaneously breaks $C_4$ rotational symmetry down to $C_2$.  As shown in Ref. \onlinecite{nie}, this has a profound effect on the phase diagram in the presence of weak disorder -- while no true incommensurate CDW order exists, the spontaneous breaking of rotational symmetry persists for disorder strength less than an order 1 critical value - a genuine vestigial nematic phase with a well-defined critical temperatures is robust in this case.  

\begin{figure}[hbt]
\includegraphics[width=0.5\textwidth]{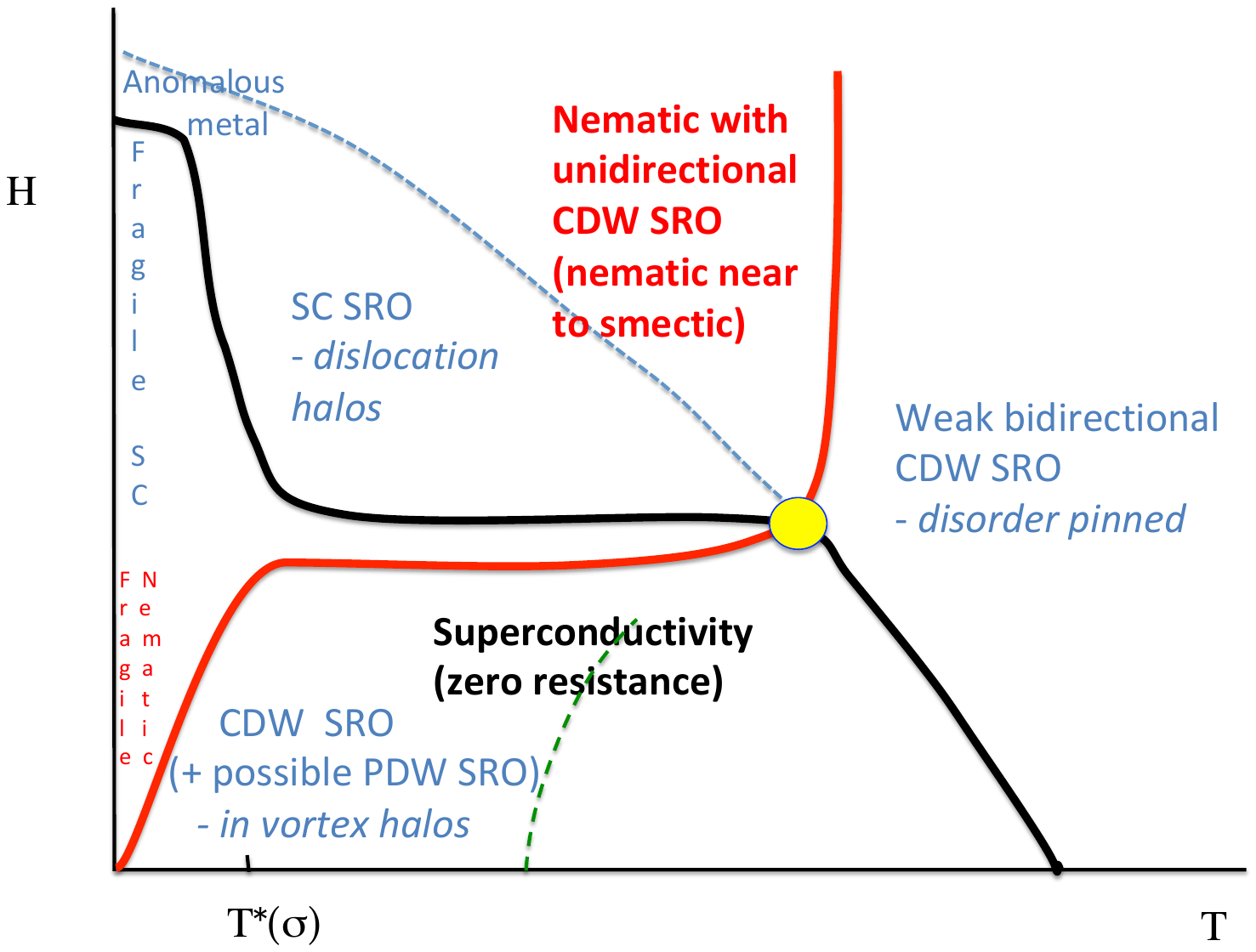}
\caption{ Schematic phase diagrams showing  competition between unidirectional CDW and SC order in a weakly disordered tetragonal system as a function of $T$ and $B$.  
The solid black line indicates the superconducting transition and the solid red line is the nematic phase boundary.  The various descriptive text are explained in the text in Sec. \ref{htc}.}
\label{Final}
\end{figure}

We thus are lead to the schematic phase diagram sketched in Fig. \ref{Final} for a putative tetragonal cuprate with weak disorder in a regime in which the competition between CDW and SC dominate the physics.  This is a decorated version of Fig. \ref{1d}.  The principle difference relative to the orthorhombic case is  that what was formally a crossover to a high field regime with substantial CDW order in Fig. \ref{1c} now appears as the thermodynamic phase boundary  of an ordered nematic phase, as in Fig. \ref{1d}.  
\footnote{Another potential new feature of the tetragonal case is that there are a new class of topological defects - domain wall defects in the Ising-nematic order parameter, across which the local orientation of the CDW order rotates.  The CDW order is generally suppressed (although it need not vanish) at these defects as well, so it is natural to expect enhanced superconducting order here as well.  
}
As before, the ``fragile SC'' is a state in which global phase coherence is mediated by the Josephson coupling between  SC halos associated with neighboring dislocations separated by a typical distance of order the CDW correlation length.  The ``fragile nematic'' is a dual of this state, in which nematic LRO is mediated by the interaction between neighboring CDW halos associated with neighboring vortices.  

There are other features we have included in Fig. \ref{Final} that go beyond the considerations of the classical effective action we have analyzed.   
In the first place, we have  conjecturally included the effect of quantum fluctuations of the SC on the phase diagram at the highest fields and lowest $T$, where they will likely give rise to an ``anomalous metallic'' regime\cite{kapitulnikrmp}.  Here, quantum fluctuations of the phase of the order parameter on each dislocation halo destroy global phase coherence, even at $T=0$, but there remain substantial SC correlations that extend across multiple halos.  
We have also indicated a regime in which PDW order is most likely to arise -- associated with vortex halos where CDW and SC order coexist.\cite{edkins-2018,PhysRevB.91.104512,PhysRevB.97.174510,PhysRevB.97.174511}

These new results potentially give some theoretical basis for an understanding of observed phenomena in various cuprates.  In particular, the fact that many features of the phase diagram in Fig. \ref{Final} correspond to observed behaviors in underdoped YBCO in a range of dopings between $0.08 < p < 0.15$, supports the conjecture that the principal features governing the phase diagram is a fierce competition\cite{PhysRevB.98.064513} between CDW and SC orders.  Some of the salient features that we have in mind are: 
\begin{itemize}
\item Recent high field transport and magnetization measurements\cite{suchitra} have revealed a phase diagram with 
a sharp upturn in the resistively determined $H_{c2}$ at temperatures $T \lesssim 2$K.  This behavior was presented as evidence of ``resilient'' SC and correlated with previous studies\cite{ong,riggs-2011,corson1999} that found evidence of SC correlations that persist to much higher fields than the typical $H_{c2}$.    The observation that the critical current is anomalously small in this low $T$ superconducting phase was adduced as the reason it had not been previously observed, but the fact that it correlates with magnetic hysteresis\cite{sebastian-2008} (vortex pinning) was interpreted as evidence that it is ``bulk'' effect.
\footnote{An obvious possibility that must be considered is that  the resilient SC reflects some form of chemical inhomogeneity in the sample that gives rise to ``filamentary'' SC.  The observed magnetic hysteresis is inconsistent with the most straightforward versions of this interpretation, but a system of structural inhomogeneities with correlations on appropriate length scales could probably provide an alternative explanation of all the observed phenomena.  There are various ways to test this, including by comparing properties of crystals with similar doped hole concentration grown under different conditions.} 
\item A form of the phase diagram that is also reminiscent of our Fig. \ref{Final} was proposed  in Ref. \cite{julien-2018} largely on the basis of NMR studies, although given a somewhat different interpretation than in Ref. \cite{suchitra}.  One difference is that the fragile SC regime inferred from NMR does not extend to as high fields as the ``resilient SC" reported in Ref. \cite{suchitra}.  More importantly, the Knight-shift was found to be approximately field independent in the regime of the phase diagram which roughly  corresponds to the region labeled ``SC SRO - dislocation halos'' in Fig. \ref{Final};  this was (sensibly) taken as evidence that SC correlations are essentially absent here.   
\item  It was conjectured in Ref. \cite{ideal}, on the basis high field X-ray diffraction data\cite{gerber-2015,chang-2016,ideal} that the ``ideal'' CDW order - {\it i.e.} the order that would arise in a disorder-free system at high enough fields to quench SC - would be unidirectional stripe CDW order with in-phase ordering ordering in the inter-plane (c-axis) direction.  However, to account for the presence of bidirectional short-range CDW correlations (with the same in-plane ordering vector) with little in the way of c-axis correlations, it was conjectured that the CDW order must be effectively inhomogeneous, with more and less ordered regions coexisting.  More recent studies of the effect of unidirectional strain on CDW order\cite{tacon-2018} are broadly consistent with this picture.  
\item Quantum oscillations with all the signatures of a non-superconducting Fermi liquid with a reconstructed Fermi surface (presumably due to the presence of the CDW) is observed\cite{doiron-leyraud-2007,sebastian-2008} in the same range of fields and temperatures that the resilient superconductivity is detected.  In common with the field independence of the Knight shift (discussed above) and of various thermal transport coefiscients\cite{taillefer-2014,julien-2018}, these observations are most readily understood under the assumption that the SC correlations are entirely quenched by the magnetic field.  
\end{itemize}

Reconciling the conflicting evidence from different measurements that appear to indicate that  the high field state is both entirely metallic and host to strong local SC correlations is difficult - conceivably impossible - in terms of any uniform electronic structure.  However, as the materials involved are highly crystalline and believed to be structurally homogeneous, there is a natural predjudice against any interpretation that invokes electronic inhomogeneities. 
\footnote{It has been established beyond dispute in STM studies of BSCCO that the low temperature electronic structure of this cuprate is highly inhomogeneous.\cite{howald-2001,lang-2002,gomes-2007}  Moreover, the dual behavior - the appearance of enhanced CDW correlations near the SC vortex cores - has been directly visualized in these materials.\cite{hoffman-2002a,edkins-2018}  By some metrics, this material is more disordered than YBCO, so a direct comparison is not possible.  However, it seems likely that what is occurring on relatively short length scales in BSCCO is likely applicable in YBCO at somewhat larger scales.}
 In this context, we reiterate that the electronic inhomogeneities we have invoked are intrinsic features of real systems which always have some degree of disorder, even when that disorder is statistically homogeneous and  the high temperature ``normal'' state shows no significant electronic inhomogeneities.  Moreover, there is an important correlation that is, in principle, experimentally testable.  More disordered regions, have stronger, but more short-range correlated CDW order, and thus will tend to have lower local values of the SC $T_c$, but higher values of $H_{c2}$.  So, for example, we would expect that light Zn doping of YBCO would produce a small decrease in the zero field $T_c$, but a broadening of the range of $T$ in which the fragile SC persists at low $T$ and large $H$.
 
 \section*{Acknowledgements}
 We thank J. Tranquada, B. Ramshaw, and P. Armitage for helpful comments and S. Sebastian for a pre-publication copy of Ref. \cite{suchitra}. This work was supported in part by the Department of Energy, Office of Basic Energy Sciences, Division of Materials Sciences and Engineering, under Contract No. DE-AC02-76SF00515.
 
  \appendix
 \section{More Concerning the Variational solution}
To obtain the phase boundary of superconductivity in the CDW ordered phase in the absence of disorder, we need to extend the variational treatment to the interior of the broken symmetry phase.  We thus consider the following non-replicated trial Hamiltonian density
\begin{equation}
\begin{split}
&{\cal H}_{tr}=
\frac{\kappa_\phi}{2}|\bm{\nabla}\phi
|^2+\frac{\mu_\phi}{8}[{\phi}^\star + \phi]
^2-\frac{m}{2}[{\phi}^\star + \phi]
\\
&+\frac{\kappa_\Delta}{2}|(-i\bm{\nabla}-2e\bm{A})\Delta
|^2+\frac{\mu_\Delta}{2}
|\Delta
|^2
\ . \\
\end{split}
\end{equation}
We have chosen a convention such that the expectation value of $\phi$ is real, in which the small amplitude fluctuations of the imaginary part 
 is 
 gapless (i.e. is the Goldstone mode). 
Then, if we express $\phi$ in terms of the real and imaginary parts, $\phi \equiv \langle \phi \rangle + \psi_r+i\psi_i$,  the mean values of various quantities are
\begin{equation}
\begin{split}
&
\langle \phi\rangle 
=\frac{m}{\mu_\phi}
\\
&\langle{[\psi_r]^2}
\rangle= T\int\frac{d^3k}{(2\pi)^3}{\frac{1}{\mu_\phi+\kappa_\phi{k_\parallel^2}+2J_\phi\cos{k_\perp}}}\\
&
\langle{[\psi_i]^2}
\rangle=T\int\frac{d^3k}{(2\pi)^3}{\frac{1}{\kappa_\phi{k_\parallel^2}+2J_\phi\cos{k_\perp}}}
\end{split}
\end{equation}
while $\langle |\Delta|^2\rangle$ is still given by Eq.\ref{delta2b}.

The self-consistency for the variational parameters, $\mu_\Delta$, $\mu_\phi$ and $m$ are
\begin{equation}
\begin{split}
&
{m^2}
={\mu_\phi^2}\Big[\alpha_\phi-\langle{\psi}_{r}^2\rangle-\langle{\psi}_{i}^2\rangle-\lambda\langle|{\Delta}
|^2\rangle\Big]\\
&\mu_\phi=-\alpha_\phi+\frac{m^2}{\mu_\phi^2}+3\langle{\psi}_{1,r}^2\rangle+\langle{\psi}_{i}^2\rangle-\lambda\langle|{\Delta}|^2\rangle\\
&\mu_\Delta=-\alpha_\Delta+2\langle
|{\Delta}|^2\rangle+\gamma\left[\frac{m^2}{\mu_\phi^2}+\langle{\psi}_{r}^2\rangle+\langle{\psi}_{i}^2\rangle\right]
\end{split}
\end{equation}
 The value of $T_c$ is extracted from the self-consistency equations as the point at which $\mu_\Delta \to 2J_\Delta -2e\kappa_\Delta B$. This set of equations are valid so long as 
 there is CDW LRO ($m\neq{0}$) but no SC LRO.
 
This extension of the variational approach is necessary to compute the phase boundary ($H_{c2}$ at low $T$) between the phase with only CDW order and the phase with coexisting CDW and SC order.  The phase boundary could be identified as the point at which $\mu_\Delta \to 2J_\Delta -2e\kappa_\Delta B$.  It typically occurs that near the putative multicritical point, the transition becomes (presumably unphysically) weakly first order.  Here, phase boundaries must be determined by comparing the variational free energy of the CDW ordered non SC phase  with that of the SC non-CDW and the fully disordered phase.  

Another artifact of the variational approach is that it predicts that all transition temperatures vanish for $J_\Delta=J_\phi =0$, i.e. it fails to incorporate the physics of the BKT transitions to phases with quasi-long-range order.  
When comparing our variational results with Monte Carlo result, we choose $J_\Delta=J_\phi=0.3$. 

\bibliography{fragilebib}

\end{document}